\documentclass[aps,prb,twocolumn,floatfix,superscriptaddress,longbibliography]{revtex4-2}
\usepackage{color}
\usepackage{amssymb}
\usepackage{amsmath}
\usepackage{bm}
\usepackage{graphicx}
\usepackage{amsmath,amssymb,bm,epsfig,color}
\usepackage[table]{xcolor}
\usepackage{siunitx}
\usepackage[color=green!60,textsize=small]{todonotes}
\usepackage{enumitem}
\usepackage[normalem]{ulem}

\newcommand{\tnabla}{\tilde{\nabla}}

\newcommand{\To}{\rightarrow}

\begin{document}

\title{Spinodal de-wetting  
 of light liquids  on graphene}
\author{Juan M. Vanegas} 
\affiliation{Department of Physics, University of  Vermont, Burlington, VT 05405}
\author{David  Peterson} 
\affiliation{Department of Physics, University of  Vermont, Burlington, VT 05405}
\author{Taras I. Lakoba}
\affiliation{Department of Mathematics and Statistics, University of  Vermont, Burlington, VT 05405}
\author{Valeri N. Kotov}
\affiliation{Department of Physics, University of  Vermont, Burlington, VT 05405}


\begin{abstract}
We demonstrate theoretically the possibility of spinodal de-wetting in heterostructures made of light--atom liquids (hydrogen, helium, and nitrogen) deposited on suspended graphene. Extending our theory of film growth on two-dimensional  materials to include analysis of surface instabilities via the hydrodynamic Cahn--Hilliard-type equation, we characterize in detail the
spatial and temporal scales of the
resulting spinodal de-wetting patterns. Both linear stability analysis and 
direct numerical simulations
of the surface hydrodynamics show micron-sized (generally material dependent) patterns of ``dry'' regions. The physical reason for the development of such instabilities on graphene can be traced back to the inherently weak van der Waals interactions between atomically thin materials and atoms in the liquid.
%
Thus  two-dimensional materials  could represent a new theoretical and technological platform for studies of spinodal de-wetting.
\end{abstract}

\maketitle

\section{Introduction}
\label{sec:intro}

One of the greatest developments in condensed matter physics in 
the last two decades
has been 
the discovery of novel two-dimensional (2D), atomically thin materials, such as graphene \cite{Antonio}.
Numerous 2D materials  structurally similar to graphene also exist, for example  the large family of
 transition-metal dichalcogenides
(e.g., MoS$_{2}$).  These can form the building blocks of the so-called
VDW heterostructures  \cite{vd1, vdwhetero}.
Van der Waals (VDW) forces control a wide variety of phenomena in nature
as they represent interactions between neutral bodies. Such interactions depend on the polarizability
of individual atoms  and materials and therefore are sensitive to the geometry and 
screening of the Coulomb force which is ultimately responsible for the VDW interaction \cite{Israelachvili}.
VDW interactions  can play an especially important role near surfaces where they control
wetting phenomena of liquids deposited on materials, contact angles, as well as pattern formation instabilities,
such as spinodal de-wetting \cite{deGennes, Bonn:2009ha, Craster, Oron}. 

\begin{figure}[ht]
\begin{center}
\includegraphics[width=0.95\columnwidth]{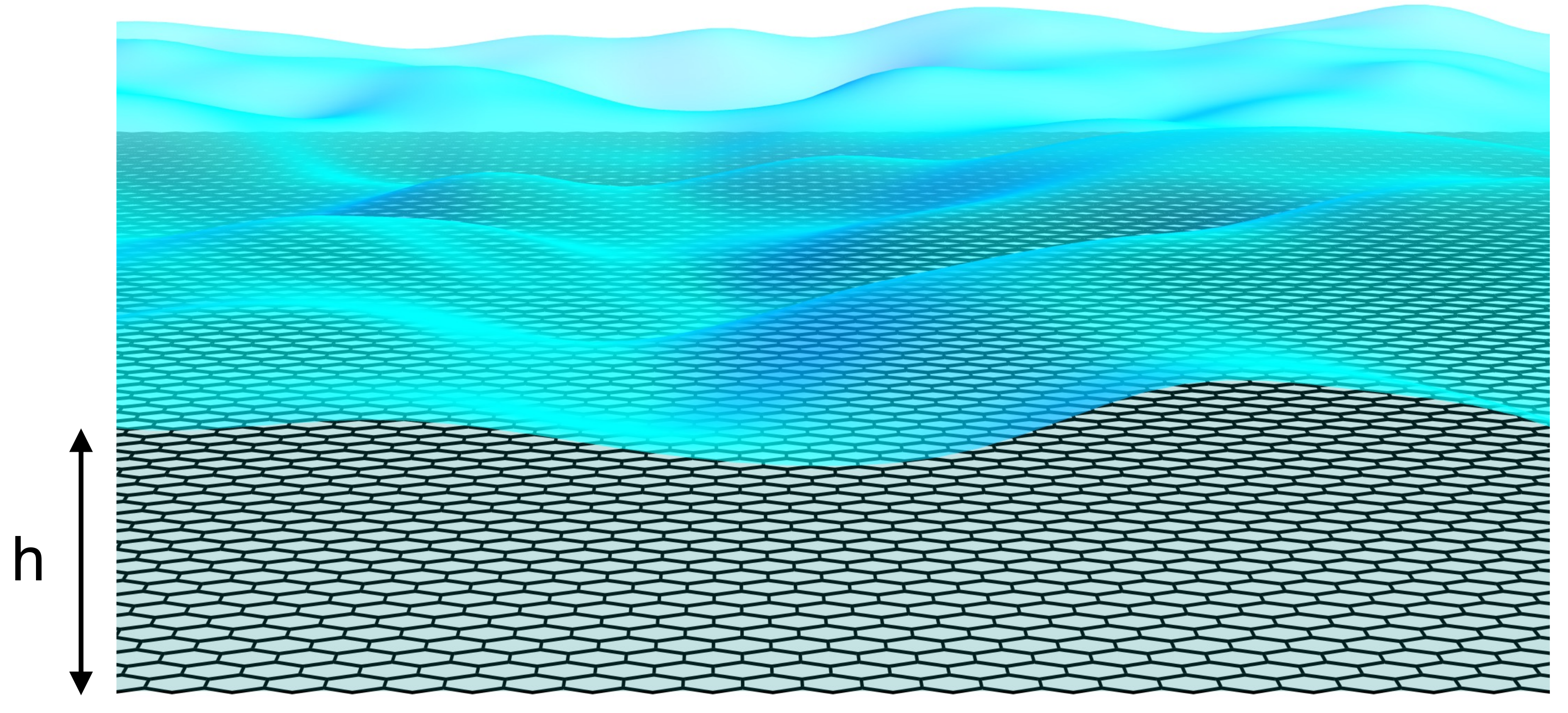}
\end{center}
\caption{Liquid film  of thickness $h$ formed on suspended graphene.}
\label{fig:wetting}
\end{figure}

 There are several important features of 2D materials that 
make them uniquely attractive candidates for studies of liquid adsorption, wetting and related VDW-driven
 phenomena. (1) First, the polarization function of 2D materials can be calculated with great accuracy. This  
 in turn leads to an excellent description of VDW forces. 
 (2)  Moreover, the polarization of graphene reflects its  
 characteristic Dirac-like electronic dispersion which can be  
 controlled
 by external factors such as application of mechanical  strain \cite{Anand,nathan, maria},
 change in the chemical potential (addition of carriers) \cite{Antonio, Kotov}, change in the dielectric environment
  (i.e., presence of a dielectric substrate affecting screening), etc. This means that VDW-related properties can be
in principle  effectively manipulated. 
(3) Also, being purely 2D structures, materials like graphene can be engineered and arranged in  various configurations. 
For example, the authors of Ref.~\cite{wetting} considered three configurations,
where the graphene sheet is either suspended in vacuum, or is submerged
in the liquid, or rests on a bulk substrate.
From the point of view of the present work, the possibility to have 
the suspended configuration,
shown in Fig.~\ref{fig:wetting},
is  the most important one.
 In this configuration, where only a single sheet of (carbon) atoms
 exerts a relatively weak VDW force on the atoms of the film, spinodal de-wetting patterns can form on the film surface.
It is important to note that  despite having atomic thickness,
graphene  is known to be  generally impermeable even  to small atoms \cite{Sun2020,Berry2013,Nair2012,Bunch2008};
hence the liquid film will remain on that side of the sheet where
it was initially present.

\begin{figure}[t]
\begin{center}
\includegraphics[width=1.0\columnwidth]{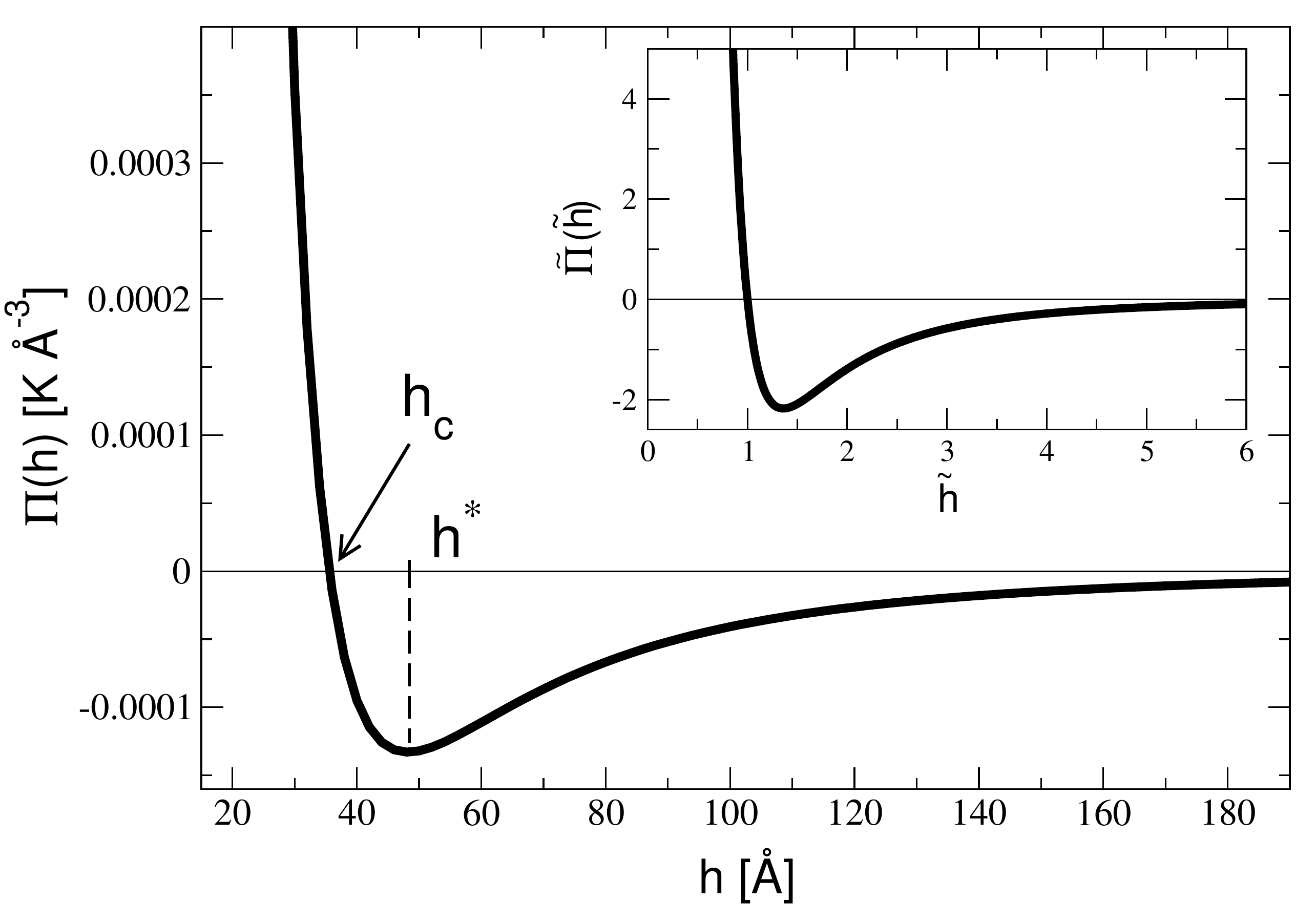}
\end{center}
\caption{Disjoining pressure $\Pi(h)$ for N$_2$. Inset:  Dimensionless disjoining pressure
as defined in the text, 
$\tilde{\Pi}(\tilde{h})=\frac{\gamma^2}{\tilde{h}^3} \left (\frac{1-\tilde{h}}{\beta + \tilde{h}} \right )$,
 where $\tilde{h} = h/a$, and the notation $a\equiv h_c$ is 
 defined in \eqref{eq:hc}.
 The parameter
 $\beta=0.37$, appropriate for N$_2$,
 and $\gamma = 5.13$ (see text).
 }
\label{fig:dp}
\end{figure}

The purpose of the present work is to study in detail 
the main characteristic scales
of the surface spinodal patterns for
three light elements: He,  H$_2$, and N$_2$, forming a liquid layer on top of suspended graphene.
The phenomenon of spinodal de-wetting itself has a long history 
\cite{Jain} (prior to that, a mathematically equivalent
analysis of spontaneous rupture of a free film was done in
\cite{Vrij}) and 
has been theoretically
predicted and detected in numerous situations involving polymers, liquid metals, etc.;
see, e.g., 
\cite{Sharma, Reiter, Mitlin, Oron, Xie, alizadeh2018, Bonn:2009ha, Mitlin2, Gentili,Rauscher2008, Seemann2001_1, Seemann2001_2,Seemann_2005, Thiele_2007Review}.
This type of de-wetting and the corresponding description bears much conceptual and technical
similarity to spinodal decomposition which describes phase separation, commonly modeled
via the Cahn--Hilliard equation (CHE)  \cite{Cahn}.
The main equation governing the 
evolution of the film thickness,
 that describes spinodal decomposition 
(the analog of the CHE in this case),
 appears in the original literature \cite{Vrij}. We will use the formulation \cite{Mitlin} which adopts the notion
 of disjoining pressure $\Pi(h)$,
 where $h$ is the (local) thickness of the liquid film.
 The shape of $\Pi(h)$, which is representative for all cases
 considered in this work, is shown in Fig.~\ref{fig:dp}.
For those values $h$ where this graph has a positive slope, an instability of the film's flat surface to 
 small fluctuations is favored, which  eventually leads to the film breakup and formation
 of spinodal de-wetting patterns. 
 A feature of the graph $\Pi(h)$ that guarantees the existence of the
 region with $d\Pi(h)/dh >0$ (for $h>h^*$; see Fig.~\ref{fig:dp})
 is the existence of a critical value $h_c$ where $\Pi(h)$ changes sign.

In order to calculate $\Pi(h)$, we rely on a previous work \cite{wetting},
where we present a detailed description
of the gaphene--liquid--vapor configuration. 
(The  analysis of that work is also applicable
to any atomically thin 2D material with liquid on top.)
It is based on the 
Dzyaloshinskii--Lifshitz--Pitaevskii (DLP) theory 
\cite{Dzyaloshinskii:1961vc, LL9},  which is the standard many-body approach for  VDW forces 
in a three-layer (substrate--liquid--vapor) configuration with given dielectric functions. 
This approach provides a very reliable description, well verified
by experiment for different substrates and liquids \cite{Israelachvili, Krim}. The work  \cite{wetting}  extends and modifies the original DLP approach, 
designed for bulk materials, to the case of 2D substrates such as graphene.
For the suspended configuration in Fig.~\ref{fig:wetting}, it was noted in \cite{wetting}  
that $\Pi(h)$  goes through zero at $h_c$ and  $d\Pi(h)/dh > 0$ when $h>h^*$, for practically all 2D materials and atoms studied there. The values of $h_c$ and $h^*$ depend strongly
on the type of liquid and  2D material substrate, but the existence of 
a region with $d\Pi(h)/dh>0$ and hence, 
an instability of the film, appears to be generic to the suspended configuration. 
This should be contrasted with the case where a thin film of a light element 
(e.g., helium) is placed over most bulk materials (e.g., graphite) \cite{1988_ChengCole},
as well as with the case of the other
two configurations of the graphene sheet considered in \cite{wetting} (i.e., where 
the sheet is either on a bulk substrate or submerged). In both of those cases,
the attractive force of atoms inside the film to the substrate is greater
than that between atoms inside the film, resulting in wetting behavior  and
stable film growth. 
Thus, the focus of the present work is on studying characteristic scales of
spinodal de-wetting patterns over a suspended graphene sheet, which represents a unique
configuration where  an instability is guaranteed to occur.

The rest of the paper is organized as follows. In Section \ref{sec:disjoining} we present results for 
the disjoining  pressure  for  three types of light liquids on graphene.
In Section \ref{sec:spinodal} we analyze the surface hydrodynamics equation (CHE) and 
present results for the  characteristic spinodal scales in the linear stability approximation.
 In Section \ref{sec:patterns} we 
 present brief details of the numerical method used
 to simulate the CHE and then 
 provide
 a detailed description of the spinodal de-wetting pattern formation and evolution.
Section \ref{sec:conclusions} contains our conclusions. In Appendix \ref{details}
we present  details of the  disjoining pressure calculation.

%
%
\section{Disjoining pressure for light liquids on graphene}
\label{sec:disjoining}

Our starting point is the analysis of Ref.~\cite{wetting},  where 
the VDW interaction energy of the configuration in  Fig.~\ref{fig:wetting} was calculated. We consider three types of light atoms: He, H$_2$ and N$_2$.
The energy is very sensitive to the atomic parameters, most notably 
the atomic polarizabilities, which are known quite accurately. The dynamical polarization of graphene is
also well known  and is an important ingredient of the calculation.
For the purpose of studying the spinodal instability, it is convenient to introduce the disjoining pressure
$\Pi(h)$, which is related to the derivative of the VDW energy as summarized in Appendix \ref{details}.

The form of $\Pi(h)$ is an important ingredient for all subsequent calculations. Based on our previous results
 \cite{wetting}, 
 the function  $\Pi(h)$ can 
be parametrized with high accuracy in the following way:
\begin{equation}
\Pi(h) = \left \{ -|\Pi_0| + \frac{\Pi_1}{h+L} \right \}\frac{1}{h^3} = \frac{|\Pi_0| }{h^3}
\left ( \frac{h_c-h}{h+L} \right ).
\label{eq:dp}
\end{equation} 
The 
film thickness $h_c$
where $\Pi(h)$ changes sign, 
which from now we label as $a \equiv h_c$, depends
on the parameters in the first part of the equation in the following way:
\begin{equation}
h_c = \frac{\Pi_1}{|\Pi_0|}  - L \equiv a .
\label{eq:hc}
\end{equation} 
The crossover length $L$ is the characteristic length-scale which separates the $-1/h^3$ and
$1/h^4$ behavior of $\Pi(h)$. 
As emphasized in \cite{wetting} and Appendix A, the existence of such a crossover  
is due to the fact that the dynamical polarization of
graphene has a very strong momentum dependence, 
reflecting the motion of Dirac quasiparticles in the layer.
The parametrization, Eq.~\eqref{eq:dp} is convenient because it reflects the presence of two physically different parts with different
signs: (a)  the $\Pi_0$ term originates from the VDW interactions with the liquid itself (thus leading to negative pressure and  tendency towards instability),
and (b) the $\Pi_1$ term comes from the graphene--liquid interaction  which favors positive pressure (and thus  stable film growth).
These two terms are written explicitly in terms of the VDW energies in Appendix \ref{details}.
 We also note that the derivation of $\Pi(h)$ 
was performed in the continuum limit, i.e. 
is valid
 for distances $h$ much larger than  graphene's lattice spacing ($\sim  1 \ \text{\AA}$). In practice this means that such 
 VDW calculations are typically used for  $h \gtrsim 20 \ \text{\AA}$ where the corrugation of the surface  is not important 
 \cite{wetting,nathan}.

Our fits for the  values of the relevant parameters for the three types of atoms, as explained in Appendix \ref{details}, lead to the following results:
\begin{eqnarray}
\text{N}_2:  \ \ |\Pi_0| &=& 72.8 \ \text{K}, \  \Pi_1 = 3592  \ \text{K\AA},  \
   L = 13.3 \ \text{\AA}, \nonumber \\
 &\Rightarrow& a = 36 \ \text{\AA}
 \label{eq:potN2}
\end{eqnarray}
\begin{eqnarray}
\text{H}_2: \ \ |\Pi_0| &=& 14.5 \ \text{K}, \  \Pi_1 =  1901  \ \text{K\AA}, \  L = 18.0 \ \text{\AA}, 
\nonumber \\
&\Rightarrow& a =114 \ \text{\AA}
\label{eq:potH2}
\end{eqnarray}
\begin{eqnarray}
\text{He}: \ \ |\Pi_0| &=& 2.09 \ \text{K},  \  \Pi_1 = 676  \ \text{K\AA}, \  L =  22.1 \ \text{\AA}, 
\nonumber \\
&\Rightarrow& a = 301 \ \text{\AA}
\label{eq:potHe}
\end{eqnarray}
A representative plot  of $\Pi(h)$ for N$_2$ is shown in Fig.~\ref{fig:dp}. 
The minimum  of  $\Pi(h)$ occurs at a distance which we label 
as: 
 \begin{equation}
 h^* = (2a-L + \sqrt{(2a-L)^2+9La})/3.
 \label{eq:hstar1}
 \end{equation}
 It is worth noticing that the values 
 of the critical distance $a$  (as well as $h^*$) are quite different for the three elements.
 Armed with the precise form of   $\Pi(h)$,  Eq.~\eqref{eq:dp}, we proceed to study 
   spinodal de-wetting pattern formation.

\section{Surface hydrodynamics: Cahn--Hilliard Equation and Spinodal  de-wetting instability}
\label{sec:spinodal}

In this section we discuss the main equations  of the theory and the linear stability analysis, appropriate
for small initial perturbations of the surface. These are compared to numerical simulations based on the finite element method which provide a complete solution and describe the full evolution in space and time.

\subsection{Main Equations}
\label{subsec:eqs}
The equation describing the evolution of $h$ has the form \cite{Mitlin, Oron, Vrij}:
 \begin{equation}
\partial_{t} h = \bm{\nabla} \cdot  \left \{ \frac{h^3}{3 \eta} \bm{\nabla} \left ( -\sigma \Delta h - \Pi(h) \right ) \right \}.
\label{eq:CHE}
\end{equation}
This is the 2D analog of the CHE, which describes bulk phase separation. 
We use the standard notation: 
 \begin{equation}
   h = h(x,y,t), \  \
 \bm{\nabla} = \ (\partial_{x}, \partial_{y}),  \    \Delta = \bm{\nabla}^2 = \partial_{x}^2 + \partial_{y}^2 .
\end{equation}
 Here $(x,y)$ is the in-plane coordinate,  $\eta$ is the liquid viscosity and $\sigma$ is the surface tension 
 (between the liquid and its vapor).
  The first term 
 on the right hand side of \eqref{eq:CHE}
 describes the resistance of the system to change of curvature (due to the Laplace pressure)  and the second term is due to the disjoining pressure.
 This equation was derived in the assumption that there is no slippage between the film and the underlying substarate (i.e., graphene in this case).
We estimated that the contribution of gravity to the evolution
of films of sub-micron thickness, considered below, is negligible.

It is convenient to re-write the equation in dimensionless coordinates. 
First we observe that the following two dimensionless combinations can be constructed naturally
\begin{equation}
\alpha \equiv \frac{|\Pi_0|}{a^2 \sigma}, \  \  \beta \equiv L/a \ .
\label{eq:abconstants}
\end{equation}
Next, we choose to measure the height $h$ in units of the critical value $a$ and 
introduce new length and time scales $\xi, \tau$. The dimensionless height and 
space/time coordinates will be denoted by tilde:
\begin{equation}
\tilde{h}= h/a,  \ \ \tilde{x}= x/\xi,  \ \ \tilde{y}= y/\xi, \ \ \tilde{t}= t/\tau .
\label{eq:dimensionless}
\end{equation}
By substituting this form into the main equation we find that we can choose:
\begin{equation}
  \xi = \frac{a \gamma}{\sqrt{\alpha}}, \   \   \    \tau = \frac{3 \eta a}{\sigma}
 \frac{\gamma^4}{\alpha^2}, 
 \label{eq:newscales}
 \end{equation}
 where $\gamma$ is an arbitrary constant and will be commented
 on below.
  With these choices the original Eq.~\eqref{eq:CHE} becomes:
\begin{equation}
\partial_{\tilde{t}} \tilde{h} = \bm{\tnabla} \cdot  \left \{\tilde{h}^3 \bm{\tnabla} \left ( -\tilde{\Delta} \tilde{h} -  
\tilde{\Pi}(\tilde{h}) 
\right ) \right \},  \  \
\tilde{\Pi}(\tilde{h}) \equiv
\frac{\gamma^2}{\tilde{h}^3} \left (\frac{1-\tilde{h}}{\beta + \tilde{h}}\right ),
\label{eq:newCHE}
\end{equation}
where $\bm{\tnabla} = \ (\partial_{\tilde{x}}, \partial_{\tilde{y}})$, $\tilde{\Delta} = \bm{\tnabla}^2$.
%
 A plot of  $\tilde{\Pi}(\tilde{h})$ is shown in the inset to
 Fig.~\ref{fig:dp}. By construction,  $\tilde{\Pi}(\tilde{h})$ changes
 sign at 
 $\tilde{h} = 1$.



\begin{figure}[ht]
\begin{center}
\includegraphics[width=1.0\columnwidth]{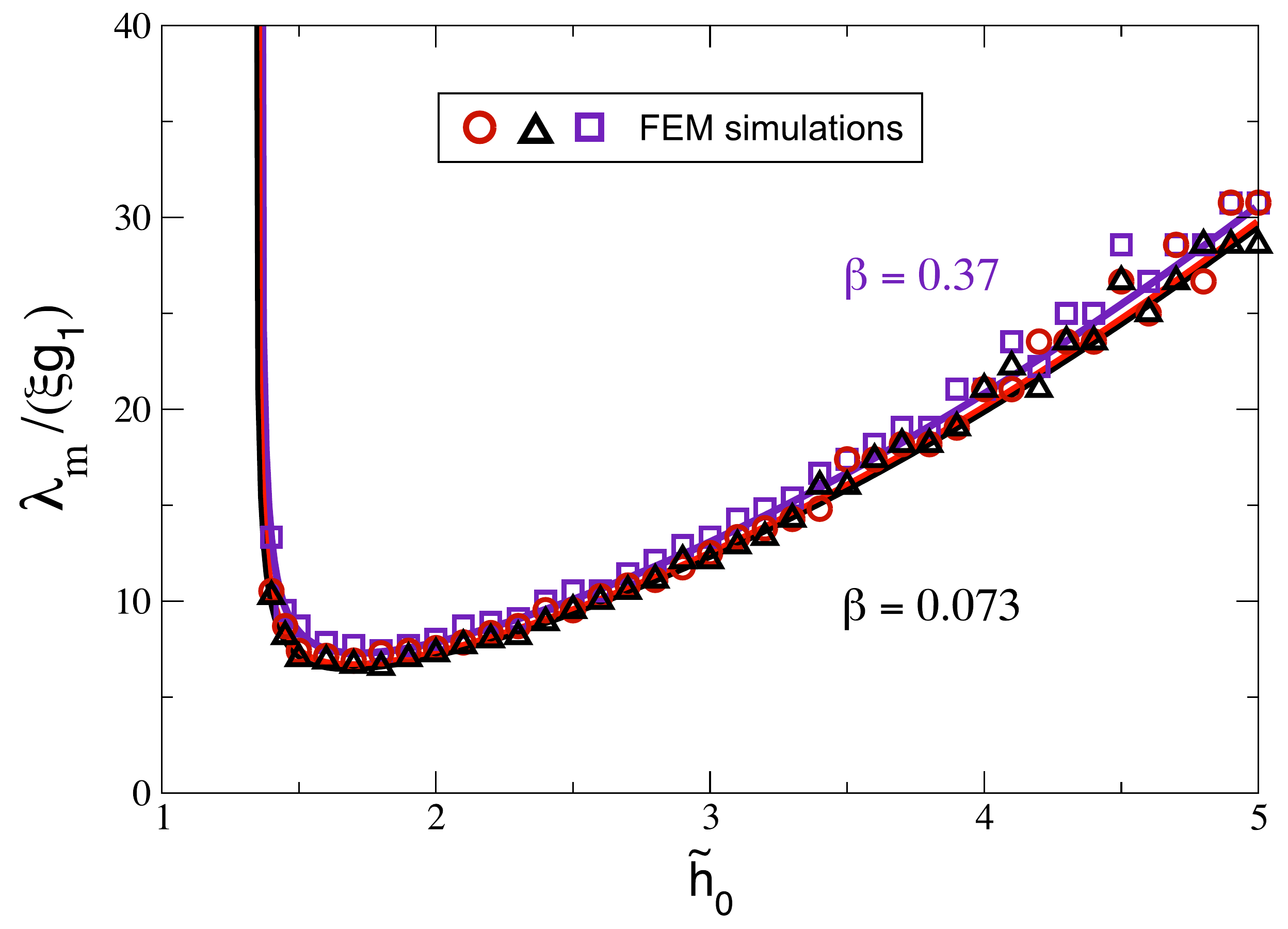}
\end{center}
\caption{The dimensionless 
spinodal wavelength $\tilde{\lambda}_m =\lambda_m /\xi$,
normalized to $g_1$ (see \eqref{eq:lamandtaum}) versus the dimensionless uniform film height 
$\tilde{h}_0=h_0/a$. For  $\beta = 0.37$ (blue), the value for N$_2$, the
onset of instability is at $\tilde{h}^*= 1.359$.  
For  $\beta = 0.16$ (red), the value for H$_2$, the
onset of instability is at $\tilde{h}^*= 1.346$.
For  $\beta = 0.073$ (black), the value for He, the
onset of instability is at $\tilde{h}^*= 1.339$. Symbols correspond to values obtained from FEM simulations 
(Section \ref{subsec:numerical}).
}
\label{fig:wl}
\end{figure}
%

\subsection{Summary of Parameters for N$_2$, H$_2$, He}
\label{subsec:parameters}

In the following sections, we characterize the short and long time scale behavior of the CHE through linear stability analysis and numerical simulations. Therefore, we summarize here the relevant scales and physical parameters for different liquids:
\begin{eqnarray}
\text{N}_2:   \beta & = & 0.37,  \  a = 36 \ \text{\AA},  \ h^* = 49 \ \text{\AA},  \nonumber \\ 
\ \xi & = & 409 \gamma  \ \text{\AA}, \
\tau  =  4  \gamma^4  \  \mu {\rm s}
\label{eq:forNa2}
\end{eqnarray}
\begin{eqnarray}
\text{H}_2:   \beta &=&  0.16,  \  a = 114 \ \text{\AA},  \ h^* =153 \ \text{\AA},  \nonumber \\
\ \xi &=& 4116 \gamma  \ \text{\AA}, \
\tau  =  392 \gamma^4  \   \mu{\rm s}
\label{eq:forH2}
\end{eqnarray}
\begin{eqnarray}
\text{He}:   \beta &=&  0.073,  \  a = 301 \ \text{\AA},  \ h^* =403 \ \text{\AA},  \nonumber \\
\ \xi &=& 27197 \gamma  \ \text{\AA}, \
\tau  =  75.5  \gamma^4 \  {\rm ms}
\label{eq:forHe}
\end{eqnarray}
For reasons explained in Sec.~\ref{subsec:linearstability}, below we use
$\gamma \approx 5.13$, so that $\gamma^4\approx 693$.
The values of $\beta, a, h^*$ are based on \eqref{eq:potN2},\eqref{eq:potH2},\eqref{eq:potHe},\eqref{eq:hstar1},\eqref{eq:abconstants},
while $\xi, \tau$ follow from  \eqref{eq:newscales}
where the following values of the surface tension and viscosity
are taken from standard tables and  literature  found in \cite{NIST-data}. 
For N$_2$ at temperature 70 K, 
$\sigma = 10$ mN/m, \ $\eta =  220\; \mu$Pa$\cdot$s;
for H$_2$ at temperature 20 K, $\sigma = 2$ mN/m, \ 
$\eta =  13.5\; \mu$Pa$\cdot$s;
for He at temperature 2.5 K, $\sigma = 0.26$ mN/m, \  
$\eta =  3.26\; \mu$Pa$\cdot$s.
The temperatures are chosen so that a liquid phase exists.

We observe that the parameter $\beta$, which appears in Eq.~\eqref{eq:newCHE}, has 
quite different values depending on the type of liquid, although we find that the solution 
depends on $\beta$ relatively weakly. More importantly,  the relevant length and time scales
can differ by orders of magnitude. 


%
\begin{figure}[h]
\begin{center}
\includegraphics[width=1.0\columnwidth]{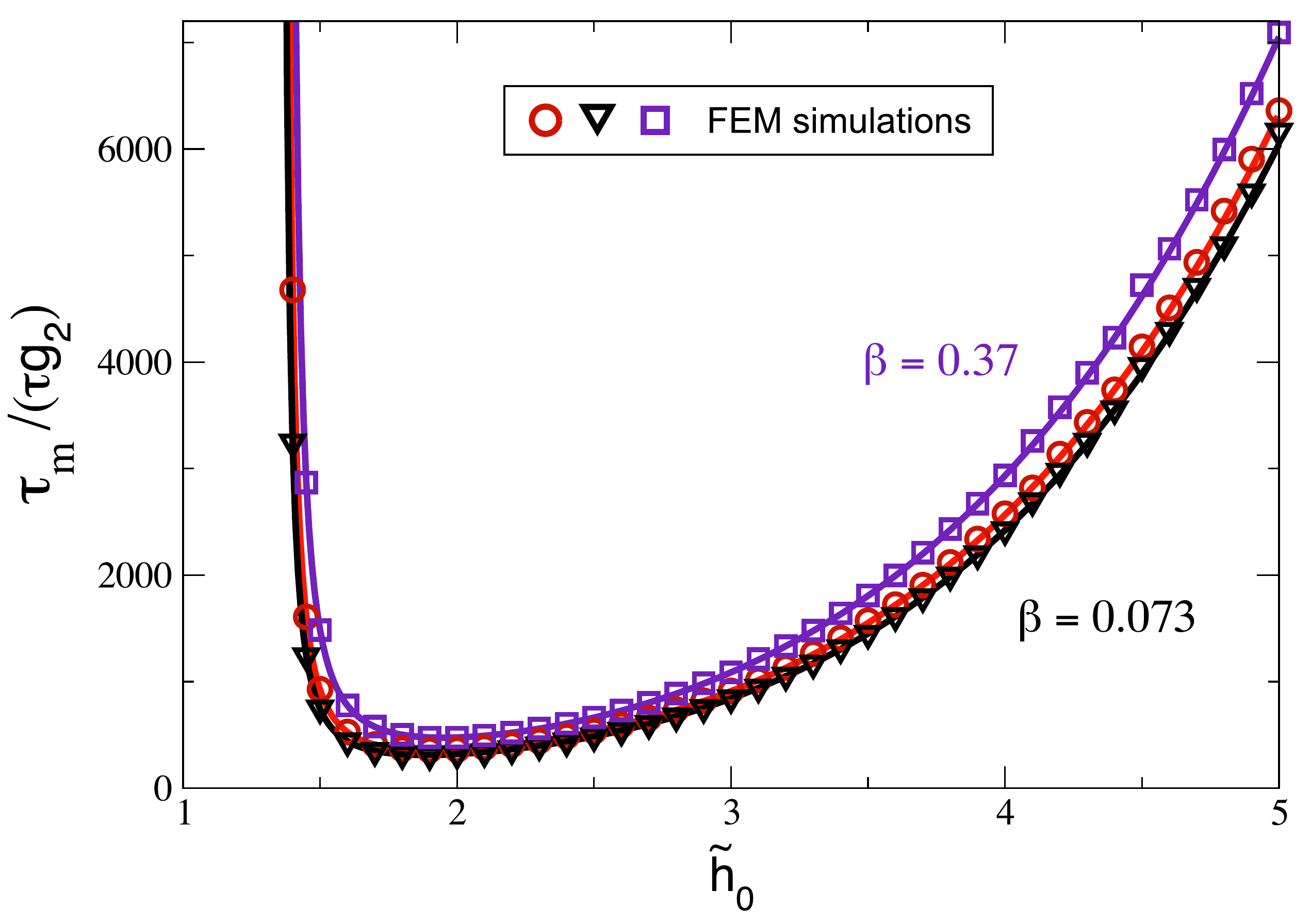}
\end{center}
\caption{The dimensionless time constant characterizing spinodal growth, 
 $\tilde{\tau}_m =\tau_m /\tau$ normalized to 
$g_2$ (see \eqref{eq:lamandtaum}), versus the dimensionless uniform film height 
$\tilde{h}_0=h_0/a$. 
The dependence on $\beta$ is 
more pronounced
compared to the wavelength $\lambda_m$. Symbols correspond to values obtained from FEM simulations.
}
\label{fig:time}
\end{figure}
%

\begin{figure*}[ht]
\begin{center}
\includegraphics[width=1.9\columnwidth]{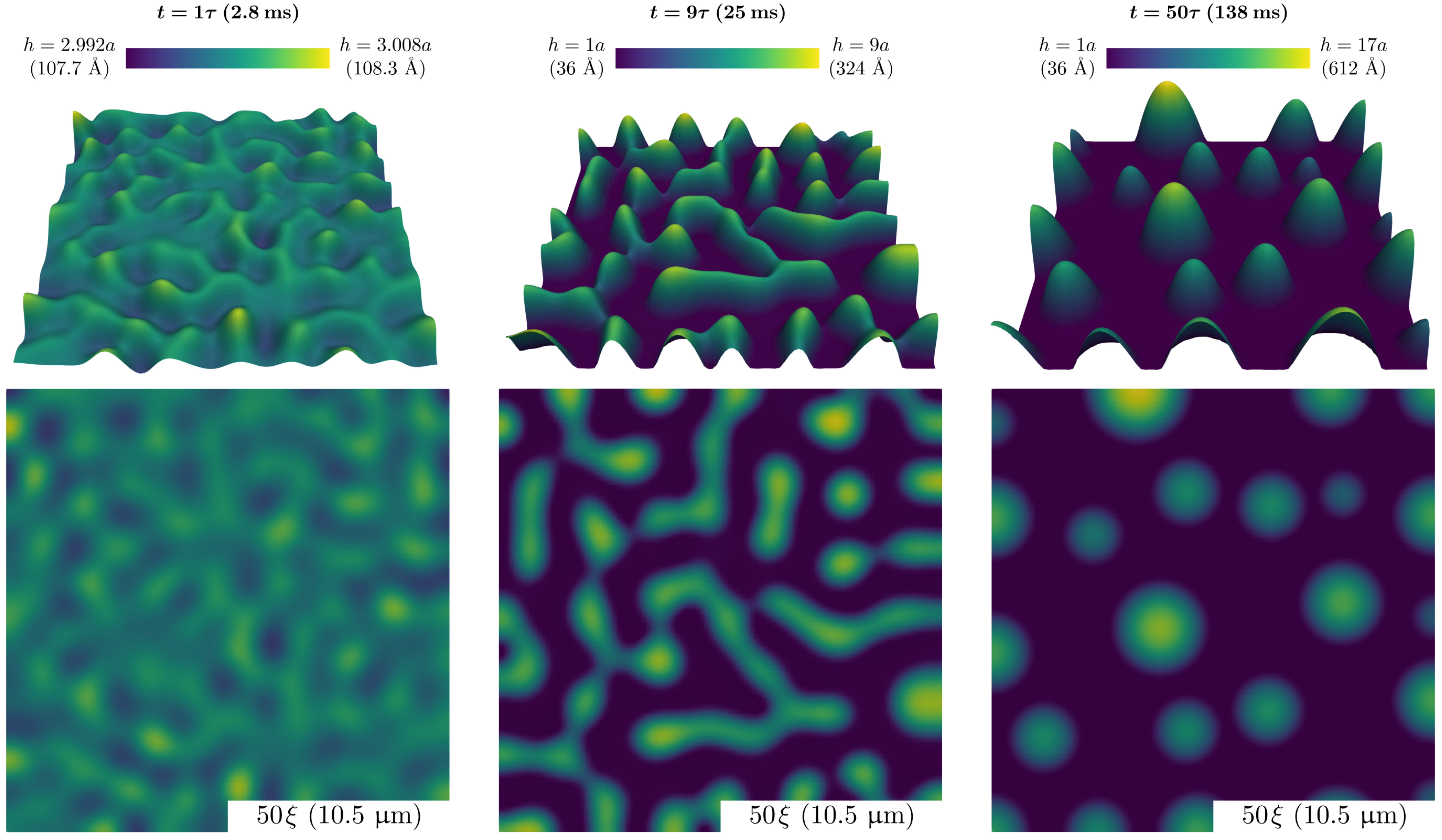}
\end{center}
\caption{Spinodal de-wetting time evolution observed in FEM simulation of N$_2$ on graphene for a liquid of initially uniform height $\tilde{h}_0=3.0$ (with $\tilde{h}^*=1.36$). During the initial stages of the simulation, the initial random 
variations
($<\tilde{h}_0\times10^{-5}$) lead to gradually increasing variations in the liquid height (left panel).
These  eventually form a well-defined spinodal de-wetting pattern (middle panel) where interconnected regions with 
excess liquid,  $\tilde{h}(x,y)>\tilde{h}^*$,
(light green/yellow colors) are surrounded by 
nearly flat regions with liquid height satisfying: $ 1 \lesssim \tilde{h}(x,y) < \tilde{h}^*$ (dark blue color). 
Note that the fact that $\tilde{h}(x,y) < \tilde{h}^*$ 
(as opposed to the two being equal)
in the background
regions is consistent with that reported in the
literature; see, e.g., \cite{Mitlin2}.
At long times, the regions of excess liquid continuously merge into distinct  ``droplets"  (localized regions of increased height).}
\label{fig:spinodal_evolution}
\end{figure*}
%

\subsection{Linear Stability Analysis}
\label{subsec:linearstability}

It is known that the spinodal decomposition (instability) regime starts at the 
value of $h$ corresponding to the minimum of $\tilde{\Pi}(\tilde{h})$ \cite{Mitlin}.
In our dimensionless notation,  
$\tilde{h}^* = h^*/a$, 
the minimum is located at 
\begin{equation}
    \tilde{h}^*= [(2- \beta) + \sqrt{(\beta-2)^2 + 9 \beta}]/3.
    \label{eq:hstar}
\end{equation}
  Thus, the instability 
  occurs for  $\tilde{h} >\tilde{h}^*$, 
  where $\tilde{\Pi}(\tilde{h})$  is negative and its derivative is positive. 
This is shown by the standard 
linear stability analysis \cite{Vrij,Jain},
as follows.

We apply a small-amplitude perturbation ($\varepsilon$) at a given wavenumber $\tilde{k}$ and imaginary frequency $\tilde{\omega}$ (both dimensionless), i.e., 
$\tilde{h}(\tilde{x},\tilde{y},\tilde{t}) =\tilde{h}_0 (1 + \varepsilon  
\ e^{i\bm{\tilde{k}}\cdot\bm{\tilde{r}}}e^{-\tilde{\omega}\tilde{t}})$,  where $\tilde{h}_0$  is the initial uniform film 
height.  By expanding 
to first order we obtain  
\begin{equation}
\tilde{\omega}(\tilde{k}) =  \tilde{h}_0^3 \tilde{k}^2 \left ( \tilde{k}^2 - \tilde{k}_c^2 \right ) ,
\label{eq:omegavsk} 
\end{equation}
where the critical wavenumber, $\tilde{k}_c$, is defined by:
\begin{equation}
\tilde{k}_c^2 \!=\! \frac{\gamma^2}{\tilde{h}_0^4} \!  \left ( \frac{\tilde{h}_0-1}{\tilde{h}_0+\beta} \right )
\! \! \left [ 3 -\frac{\tilde{h}_0(1+ \beta)}{(\tilde{h}_0-1)(\tilde{h}_0+\beta)}\right ]\! =\!   \frac{d \tilde{\Pi}(\tilde{h}_0)}{d \tilde{h}_0}.
\label{eq:kc}
\end{equation}
According to \eqref{eq:omegavsk},
an unstable mode exists as long as
$\tilde{k}_c^2>0$. From 
\eqref{eq:kc}, one can show that this occurs for 
$\tilde{h}_0>\tilde{h}^*$, where
$\tilde{h}^*$ is defined in 
\eqref{eq:hstar}. 
Thus, films thicker than 
$\tilde{h}^*$ are unstable. 
 An instability occurs
 for  wavenumbers where $\tilde{k} < \tilde{k}_c$.
The fastest growing mode is the one that has the largest  $(-\tilde{\omega})$,
which corresponds to the wavenumber $\tilde{k}_m = \tilde{k}_c/\sqrt{2}$.
This maximum instability growth rate
is $|\omega(\tilde{k}_m)| = \tilde{h}_0^3 \tilde{k}_m^4$, which leads to the  time constant  $\tilde{\tau}_m = (\tilde{h}_0^3 \tilde{k}_m^4)^{-1}$, 
meaning that the perturbation grows as $\sim e^{\tilde{t}/\tilde{\tau}_m}$.
The  spinodal wavelength (corresponding to the fastest growing mode) is 
\begin{equation}
\tilde{\lambda}_m = 2 \pi/\tilde{k}_m=2 \pi \sqrt{2}/\tilde{k}_c.    
\label{eq:lambdam}
\end{equation}

For  values 
$\tilde{h}_0 \gg \tilde{h}^*$,
one extracts the  
asymptotic behavior 
\begin{subequations}
\begin{equation}
\tilde{\lambda}_m \sim  g_1\ \tilde{h}_0^2, \  \
\tilde{\tau}_m \sim  g_2 \ \tilde{h}_0^5,  \  \ {\mbox{for}}  \  \ \tilde{h}_0 \gg 1, 
\label{eq:lamandtaum}
\end{equation}
where
\begin{equation}
    g_1 = \frac{2 \pi \sqrt{2}}{\gamma\sqrt{3}},
    \qquad
    g_2=\frac{4}{9 \gamma^4}.
\label{eq:g1andg2}
\end{equation}
\end{subequations}
The choice $\gamma = 5.13$ results in $g_1 =1$, which leads to a 
simple
form of the asymptotic dependence of the most unstable wavelength on the 
 film height (in non-dimensional units). 
We found this to be 
a convenient choice in the numerical simulations, but any other choice of $\gamma$ is also acceptable.

Plots of the  spinodal wavelength (Fig.~\ref{fig:wl}) and the  spinodal growth time constant (Fig.~\ref{fig:time}) show divergence at the instability threshold and then increase as power laws for larger film heights. At the onset of instability, i.e.,
for $\tilde{h}\To \tilde{h}^*+0$,
the critical wavenumber is $\tilde{k}_c\sim (\tilde{h} - \tilde{h}^*)^{1/2}$, and therefore the 
most unstable
wavelength
diverges as
$\tilde{\lambda}_m \sim (\tilde{h} - \tilde{h}^*)^{-1/2}$.
The time scale of the instability, $ \tilde{\tau}_m \sim (\tilde{h} - \tilde{h}^*)^{-2}$, diverges 
even
more strongly than the wavelength.

In Fig.~\ref{fig:wl} and Fig.~\ref{fig:time} we 
present these 
values $\tilde{\lambda}_m$ and $\tilde{\tau}_m$ 
along with the corresponding quantities
obtained from numerical simulations (see Section 
\ref{sec:patterns})
by calculating a radially averaged 2D Fourier transform of $\tilde{h}(x,y)$ at each time step and identifying the fastest growing modes. We find excellent agreement between the results from linear stability analysis and numerical simulations across all values of the initial heights tested. The dependence on the parameter $\beta$, which varies with the type of liquid, is relatively weak, practically non-existent for $\lambda_m$ and somewhat more pronounced for $\tau_m$.

\begin{table}[ht]
\centering
 \renewcommand{\arraystretch}{1.5}
  \setlength\tabcolsep{9pt}
    \begin{tabular}{c c c  c c}
\hline\hline
 Atom &  $a$&  $\xi$ &   $\tau$   &$\beta$\\ 
\hline
N$_2$& 36  \text{\AA} &  0.210 $\mu$m&    2.77$\times 10^{-3}$ s& 0.37 \\
H$_2$& 114  \text{\AA} &   2.11 $\mu$m&  0.271 s& 0.16\\
He   & 301  \text{\AA}&  13.9   $\mu$m& 52.3 s& 0.073 \\
\hline
\end{tabular}
\caption{Time and length scales for different elements,
computed from Eqs.~\eqref{eq:forNa2}--\eqref{eq:forHe}
for $\gamma=5.13$.
Here $a$ represents the scale of the height $h$,
and $\xi$ is the scale in the planar ($x$ and $y$) direction as introduced in Eq.~\eqref{eq:dimensionless}, and
$\tau$ is the time scale. The quantity $\beta$ is 
defined in \eqref{eq:abconstants}. 
}
\label{tab:nonlin}
\end{table}

\begin{figure}[b]
\begin{center}
\includegraphics[width=1.0\columnwidth]{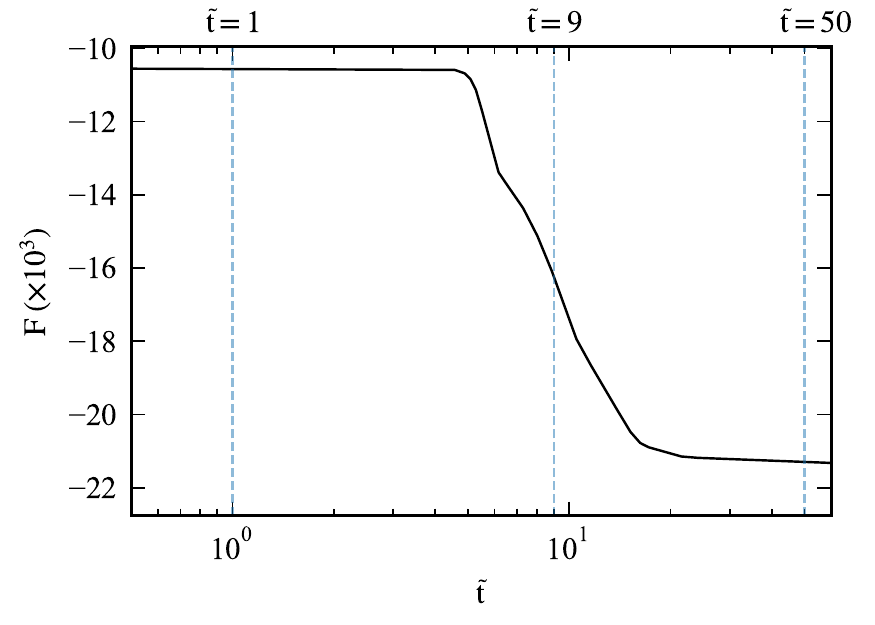}
\end{center}
\caption{
Evolution of the free energy Eq.~\eqref{eq:FE} with time
for the N$_2$ film, obtained by FEM. Total value of the free energy depends on the area of the system ($A=100\xi\times100\xi$)
}
\label{fig:energy}
\end{figure}

\begin{figure}[b]
\begin{center}
\includegraphics[width=0.95\columnwidth]{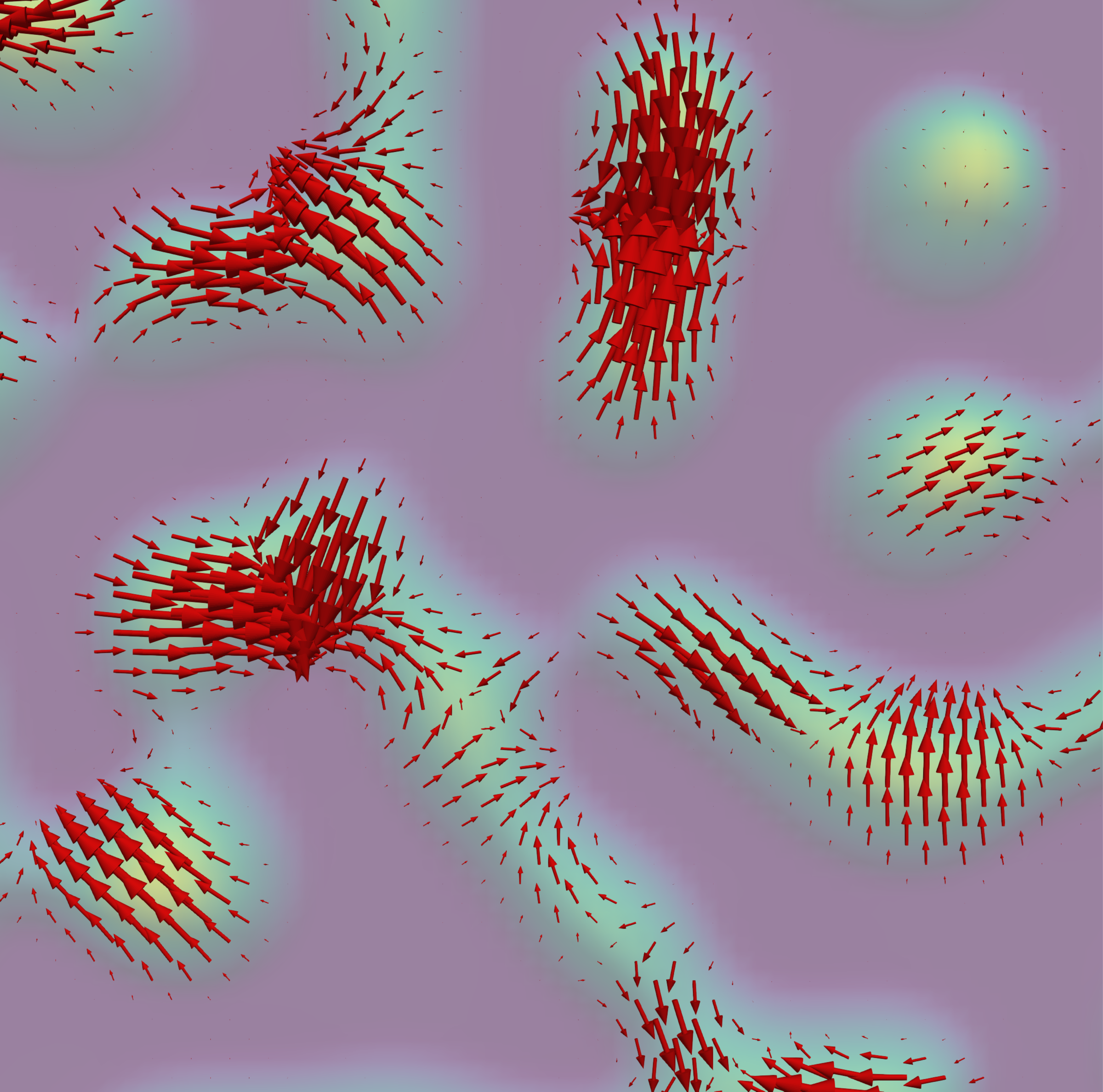}
\end{center}
\caption{Close-up of the flux vector field ($\bm{\mathrm{\tilde{J}}}$, red arrows) for the N$_2$ FEM simulation shown in Fig.~\ref{fig:spinodal_evolution} (center panel). Arrow sizes are scaled by the magnitude of the flux vector at a particular $x-y$ location. Note how the flux vector field depicts different types of motions within the fluid including the translational motion of large features as well as the merging of neighboring ones (regions with high density of arrows). Height data shown in the background where purple/dark colors correspond to values of $\tilde{h} \approx 1$ and yellow/light green colors to $\tilde{h} \leq 9$.}
\label{fig:vector_field}
\end{figure}
%

\section{Numerical simulations of spinodal de-wetting}  
\label{sec:patterns}

\begin{figure*}[t]
\begin{center}
\includegraphics[width=1.9\columnwidth]{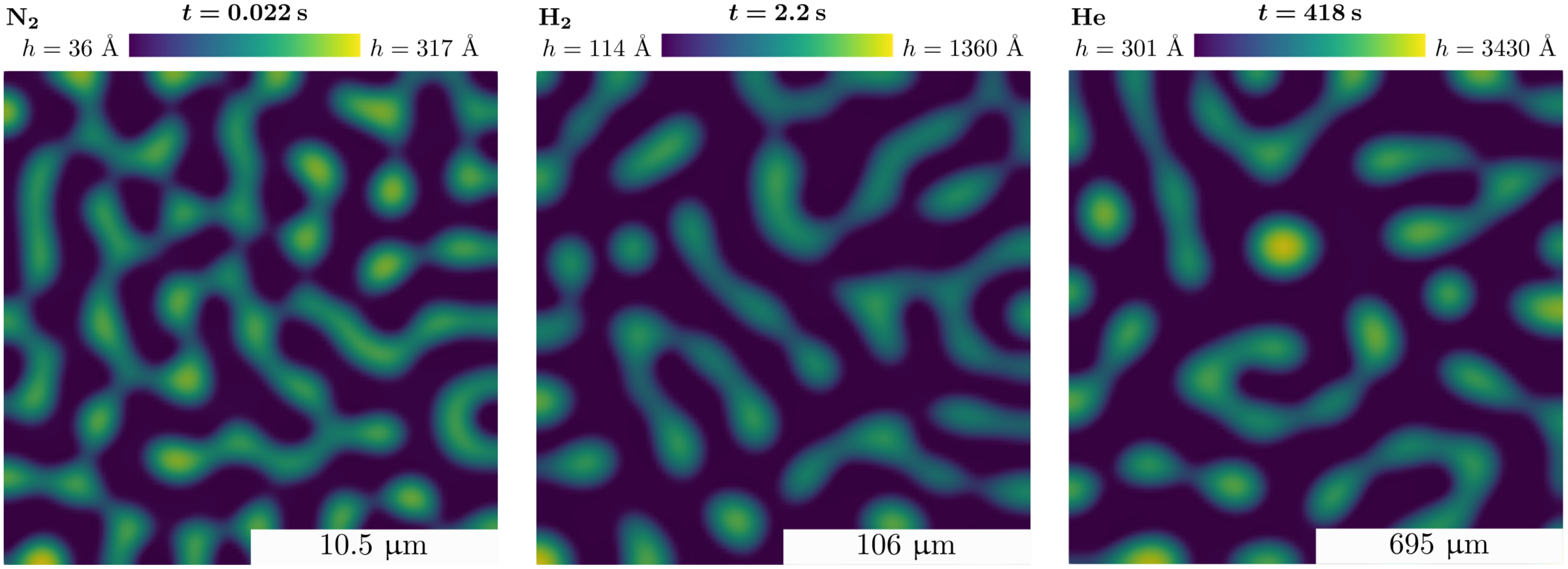}
\end{center}
\caption{Comparison of spinodal de-wetting patterns observed in FEM simulations of liquids on graphene for N$_2$ (left panel), H$_2$ (center panel), and He (right panel). For all three cases shown, the dimensionless parameters are the same ($\tilde{h}_0=3.0$, $t=8\tau$, and $100\xi$ simulation size) except for $\beta$ (see Table \ref{tab:nonlin}). While the patterns are qualitatively similar regardless of the liquid, the length and time scales are vastly different as discussed in the main text.}
\label{fig:comparison}
\end{figure*}
%

\subsection{Finite Element Simulations}
\label{subsec:numerical}

To perform numerical simulations of Eq.~\eqref{eq:newCHE}, we first rewrite it
in the form of a continuity equation:
\begin{equation}
\partial_{\tilde{t}} \tilde{h} = \bm{\tnabla} \cdot \bm{\mathrm{\tilde{J}}}
\label{eq:continuity}
\end{equation}
where
\begin{align}
\bm{\mathrm{\tilde{J}}} = \tilde{h}^3 \bm{\tnabla} \nu(\tilde{h}) 
\label{eq:flux}
\end{align}
is the dimensionless particle flux vector field, and 
we have defined for convenience the quantity $\nu(\tilde{h}) = -\Delta \tilde{h} - \tilde{\Pi}(\tilde{h})$.
To guarantee that 
the mass of the liquid over a given area of the substrate
is conserved, we impose the following zero-flux boundary conditions:
\begin{equation}
\bm{\tnabla}\nu \cdot \hat{n} = 0 \  \ 
\mbox{on $C$};
\label{eq:zerofluxBC}
\end{equation}
here curve $C$ is the boundary of the given area and 
$\hat{n}$ is the unit normal vector to the boundary. 
Indeed, integrating over the given area 
and applying the 2D version of the divergence theorem, we obtain $\partial_{\tilde{t}} \int \tilde{h} d\tilde{S} = 
\int_C \tilde{h}^3 (\bm{\tnabla}\nu \cdot \hat{n})dl =0$, where the  
last equation follows from \eqref{eq:zerofluxBC}. 
(The mass with all units restored is $n_{\rm liquid} \int h dS$, where $n_{\rm liquid}$ is the liquid density.)

Numerical simulations of 
Eqs.~\eqref{eq:continuity} and \eqref{eq:flux}
 were performed in Python with the FEniCS automated finite element method (FEM) package \cite{Fenics1,Fenics2,Dolfin}. A standard Lagrange finite-element basis was used to solve these 
 equations
 variationally \cite{Fenics2}. Time integration was performed using the standard finite difference Crank--Nicolson method \cite{Crank-Nicolson}. Sufficiently small time steps were chosen in order to facilitate convergence of the FEM solvers depending on the parameters for each species (see Table \ref{tab:nonlin}) and  film thickness values. Numerical accuracy was monitored by checking conservation of total mass at each time step (${dM}/{M_{\mathrm{total}}}<10^{-14}$).

The starting condition for all simulations corresponded to the 
spatially uniform film of thickness ($\tilde{h}_0$) with very small random 
variations
($<\tilde{h}_0\times10^{-5}$). Neumann boundary conditions were applied at the edges of the simulation box (Eq.~\eqref{eq:zerofluxBC}). Analysis of the FEM simulations was performed with the NumPy and SciPy libraries \cite{NumPy,SciPy}.

\subsection{Time Evolution of Spinodal De-wetting Patterns}
\label{subsec:numerresults}

The spinodal de-wetting patterns for N$_2$ (taken as an example) with $\tilde{h}_0=3$ obtained from numerical simulations are presented in Fig.~\ref{fig:spinodal_evolution} for three different times (corresponding to the free energy evolution in  Fig.~\ref{fig:energy}). These show the characteristic spinodal surface patterns as time increases, culminating in large height fluctuations at late times. For N$_2$ (Fig.~\ref{fig:spinodal_evolution}), the observed distance between features at the initial/intermediate stages is $\sim 1  \ \mu \mbox{m}$ in agreement with the spinodal wavelength values shown in Fig.~\ref{fig:wl} (in units of the length scale $\xi \approx 0.2  \ \mu\mbox{m}$, see Table \ref{tab:nonlin}).

The observed time evolution of the liquid film can be further characterized by considering the free energy:
\begin{equation}
F = \iint \left \{ \frac{1}{2} | \bm{\nabla} \tilde{h}|^2  +  U(\tilde{h}) \right \}d \tilde{x}d \tilde{y},
\label{eq:FE}
\end{equation}
where the potential energy  $U(\tilde{h})$ is defined as 
$\partial U/\partial {\tilde h} = - \tilde{\Pi}(\tilde h)$, 
\begin{equation}
 U(\tilde{h}) = \frac{\gamma^2}{2\beta}\frac{1}{\tilde{h}^2} - \frac{\gamma^2 (1+\beta)}{\beta^2}\frac{1}{\tilde{h}}+ \frac{\gamma^2 (1+\beta)}{\beta^3} \ln{\left ( 1 + \frac{\beta}{\tilde{h}} \right )}.
\end{equation}
The free energy in Eq.~\eqref{eq:FE} decreases with time and is constant only on stable stationary solutions if/when they exist: $\frac{dF}{dt} \leq 0$ \cite{Mitlin}. Values of the free energy for the N$_2$ numerical simulation are shown in Fig.~\ref{fig:energy}. During the initial time evolution, $\tilde{t}<5$, the small-scale fluctuations 
of the film thickness
are reflected in the approximately constant energy. At intermediate times, $5<\tilde{t}<20$, the energy rapidly changes as the spinodal fluctuations grow macroscopically and well-defined ridges of material accumulate above a 
nearly
uniform film surface of thickness  $\tilde{h}\gtrsim 1$
(see caption for Fig.~\ref{fig:spinodal_evolution}). At larger 
(dimensionless) times,
$\tilde{t}>20$, the energy enters a slowly changing regime as the ridges merge into 
isolated droplets 
that accumulate the excess liquid,
surrounded by large areas of flat surface.
The above stages of the film evolution follow a
well-established sequence, for example as reported in \cite{Sharma} for a different physical system (different $\Pi(h)$).

The redistribution of mass in the process of de-wetting can be more clearly observed with the help of the flux vector, $\bm{\mathrm{\tilde{J}}}$,  as shown in Fig.~\ref{fig:vector_field}. While the total mass is conserved, as discussed previously, there is significant flow toward regions of larger height, relative  to the uniform value.

Having examined patterns at different times for N$_2$,
we turn our attention to comparing the evolutions 
for different elements. 
The time and length scales for the three elements 
listed in Table \ref{tab:nonlin}
are quite different.
Namely, the time scale is the shortest for N$_2$ and longest
for He; this results in the evolution of He being much slower than that of the other two liquids in physical units. 
 For example, for the nondimensional  height 
$\tilde{h}_0=3$, the spinodal growth time scale for He is $\tau_m \sim 10  \ \text{s}$,
while this time scale  for H$_2$ and  N$_2$ is 
 $\sim 10^{-1}\ \text{s}$ and $10^{-3}\ \text{s}$, respectively.
Moreover, as shown in Fig.~\ref{fig:comparison}, 
the same {\em dimensionless} simulation time results in patterns corresponding  to 
somewhat 
later stages of evolution for He than for H$_2$ and N$_2$.
We hypothesize that this difference can be caused by the
different values of $\beta$ for these three elements,
because in all other respects their dimensionless equation
\eqref{eq:newCHE} is identical. 
In the same Figure, we can see that at the intermediate
stage of the evolution, the characteristic size of the 
emerging ``ridges" approximately follows the scale of
the most unstable wavelength, whose values for the
initial film thickness in question, $\tilde{h}_0=3$,
are: \ 
$\lambda_m \sim 1  \ \mu \mbox{m}$ (for N$_2$),   
$\lambda_m \sim 10  \ \mu \mbox{m}$ (for H$_2$), and 
 $\lambda_m \sim 100 \ \mu \mbox{m}$ (for He).
 Recall from \eqref{eq:lamandtaum} that $\tilde{\lambda}_m\propto \tilde{h}_0^2$. 
 We also observe  that 
 as the initial thickness $\tilde{h}_0$ increases, the 
 diameter of the droplets formed at the terminal stage of the 
 evolution also increases, albeit slower than quadratically.
  This is consistent with similar observations in a different physical context  \cite{SharmaReiter_Dropssize}.

\section{Conclusions and Outlook}
\label{sec:conclusions}
This work predicts the existence of  surface de-wetting patterns for  light liquids
on suspended graphene and 
investigates in detail the  spatial and temporal scales that characterize those patterns. 
The first important step in the problem is the 
calculation
of the disjoining pressure $\Pi(h)$,
using the approach laid out in 
Ref.~\cite{wetting}.
This function can be determined very accurately for
various elements on graphene since the atomic parameters and  graphene's polarization can
be calculated with great accuracy. In fact, the general shape shown in Fig.~\ref{fig:dp} is quite universal and
representative of   numerous  two-dimensional materials such as members of the dichalcogenides family (MoSe$_2$, MoS$_2$,  WSe$_2$, WS$_2$).
 For all of these, the film thickness $h^*$ at which spinodal de-wetting starts (for He liquid) is between $100 \ \text{\AA}$ and $300 \ \text{\AA}$ 
  \cite{wetting}. Applying additional perturbations to graphene itself, such as 
  electronic (or hole) doping via external voltage, also affects $h^*$, generally increasing it \cite{wetting}.
Therefore spinodal patterns  are possible for liquids on all of those materials as well, the main difference being in the various characteristic length and time-scales which are very material specific.
We also point out that the most important physical assumption in our analysis leading to Fig.~\ref{fig:dp} and everything that  follows is that graphene (or any of the other 2D materials) are in the suspended configuration,
since only in this case a finite $h^*$ is predicted, whereas the presence of an additional (bulk) substrate
creates too much VDW attraction and sends $h^*$ to infinity.  
The  possibility of suspended configurations is  
a unique feature
of 2D materials.

An advantage of studying light liquids, as we have done in this work,  is that their spinodal de-wetting characteristics 
can be predicted theoretically very accurately,  in the relatively  low-temperature regime where liquid phases exist.
For complex liquids, including liquid metals,  this  would not be a simple task. 
Additional real-world factors such as, for example, bending of suspended graphene sheets (or other 2D materials) should
also in principle be taken into account in the calculation of VDW interactions.

The spinodal de-wetting patterns 
observed numerically 
(see Fig.~\ref{fig:spinodal_evolution}) for various liquids on graphene
are quite universal in shape and time evolution when written in dimensionless form.
 The main difference is in the time and length scales for
different elements (Table \ref{tab:nonlin} and Figures~\ref{fig:wl}, \ref{fig:time}). 
We also found that 
the spinodal wavelengths,
and subsequently the spatial scales of the emerging patterns (ridges and droplets),
are generally quite long compared to the critical film thickness for spinodal onset (which is up to several hundred $\text{\AA}$),  and range between $1 \ \mu \mbox{m}$ and $100 \ \mu \mbox{m}$
depending on the liquid.

While in this work we considered the instability of the film surface with respect to small initial perturbations, it should be noted that different dynamics may result, for certain ranges of the initial film thickness, when the film is subject to  finite perturbations to its shape. A study of the resulting metastable and ``nucleation-dominated" regimes 
(see, e.g., \cite{2001_Thiele_PRL,2003_NatureMat,Sharma_ManyPaths,Thiele_2007Review})
of a film's surface evolution is outside the scope of this paper.

We hope this work stimulates further theoretical and experimental  research related to the physics of spinodal de-wetting on 2D atomically thin crystals,
especially since  this phenomenon appears to be a universal feature for this class of materials. We emphasize again the most important advantages of 2D materials, such as graphene:

\begin{itemize}[leftmargin=0.3cm]
\setlength\itemsep{-0.1em}
\item[$-$] The spinodal de-wetting instability is a generic phenomenon in such materials and  occurs spontaneously  
 at the instability onset $h^*$ due to the fact that 2D structures are weak adsorbers, i.e., their VDW potential is not strong enough to maintain 
a film with uniform thickness in excess of $h>h^*$. 
\item[$-$] Given that 2D material parameters are known with great accuracy, the  spinodal de-wetting onset $h^*$ and the evolution of the spinodal de-wetting patterns can be reliably predicted for liquids with well-established polarization characteristics. 

\item[$-$] Because graphene and 2D materials can be also manipulated via external factors such as carrier doping, strain, etc., this can be used as a guiding principle for creation and control of de-wetting patterns. For example 
a range of values $h^*$  was found in \cite{wetting} for graphene and other 2D materials, such as monolayer dichalcogenides, which could lead to applications in micro-pattern design \cite{Gentili}.
\end{itemize}

\begin{acknowledgments}
We are grateful to Adrian Del Maestro and Peter Taborek for numerous stimulating discussions related to the physics
of wetting and  wetting instabilities. 
J.M.V., T.I.L. and  V.N.K. gratefully acknowledge  financial support 
from NASA Grant No. 80NSSC19M0143.

\end{acknowledgments}

\appendix
\section{Details of Disjoining Pressure Calculations for Light Atoms on Graphene}
\label{details}
Here we summarize the results  of calculations related to the determination of the disjoinging pressure $\Pi(h)$, Eq.~\eqref{eq:dp}, 
which is used to extract the relevant parameters for different atoms, 
Eqs.~\eqref{eq:potN2},\eqref{eq:potH2},\eqref{eq:potHe}. The form of Eq.~\eqref{eq:dp} follows from the microscopic DLP theory \cite{Dzyaloshinskii:1961vc, LL9}, 
when applied to 2D materials, which 
describes VDW interactions in anisotropic (layered) situations such as liquids on solid substrates \cite{Israelachvili, Dzyaloshinskii:1961vc,LL9}. The standard calculations and typical applications assume a bulk (usually dielectric)
 substrate with a liquid formed on top, in equilibrium with its vapor. In \cite{wetting} one of us and collaborators extended the standard theory to  several physical situations involving 2D materials,
  and in particular to the case when a 2D semimetal,  such a graphene, is used as a substrate instead of a bulk material
  (as shown in Fig.~\ref{fig:wetting}). 
  We refer the reader to \cite{wetting} for details of calculations. The ground state energy of this system  can be written as (we set $\hbar = 1$):
\begin{equation}
U_{\mbox{vdw}}(h) = \frac{1}{(2\pi)^{3}}\int d^2{\bf q} \int_{0}^{\infty} d\omega  ({\cal{U}}_d (q,i\omega)+ {\cal{U}}_g(q,i\omega)),
\label{eq:totalvdw}
\end{equation}
where ${\cal{U}}_d (q,i\omega)$  describes the liquid with dielectric function $\varepsilon(i\omega)$
and thickness $h$, without a substrate and  with liquid vapor on top  (taken as vacuum, dielectric constant equal to one),
\begin{equation}
{\cal{U}}_d (q,i\omega)= \frac{(\varepsilon(i\omega) -1)(1 - \varepsilon(i\omega) )}{(\varepsilon(i\omega) +1)(1 + \varepsilon(i\omega) )}e^{-2qh},
\label{eq:dielectricvdw}
\end{equation}
and ${\cal{U}}_g(q,i\omega) $ is the graphene substrate--liquid interaction part: 
\begin{equation}
{\cal{U}}_g(q,i\omega)  =  \frac{ \left (\frac{-4 \pi e^2 \chi(q,i\omega)}{q  (\varepsilon(i\omega)
    +1)} \right) \left (\frac{\varepsilon(i\omega) -1}{\varepsilon(i\omega) +1}
\right ) \left (\frac{2\varepsilon(i\omega) }{\varepsilon(i\omega) +1} \right )
}{ 1 -\frac{4 \pi e^2 \chi(q,i\omega)}{q  (\varepsilon(i\omega)
+1)} } e^{-2qh}.
\label{eq:graphenevdw}
\end{equation}
Equations \eqref{eq:dielectricvdw} and \eqref{eq:graphenevdw} follow from more general expressions (describing different geometries)  derived in \cite{wetting}. 
Here $q=|\textbf{q}|$ is the magnitude of the in-plane momentum and $\chi(q, i\omega)$ is graphene's polarization
function which is known to be \cite{Kotov}:
\begin{equation}
\chi(q, i\omega) = -\frac{1}{4}
\frac{q^2}{\sqrt{v^2q^2+\omega^2}},
\label{eq:grpolarization}
\end{equation}
where $v = \SI{6.6}{\eV\angstrom}$ is the  velocity of the Dirac quasiparticles.
We have modified somewhat the notations used in \cite{wetting} in order to achieve consistency
with the symbols across the present paper.

It should be emphasized that Eq.~\eqref{eq:totalvdw} describes any 2D material (not only graphene), with a 
 dynamical polarization  $\chi(q, i\omega)$, in the suspended configuration. 
This allows one to compute the spinodal instability threshold $h^*$ 
and indeed the function $\Pi(h)$ with high accuracy.
We also note: (1) Relativistic  effects are negligible in the range of distances of interest to us  (up to hundreds of \AA) \cite{nathan,wetting}.
(2)  The energy in Eq.~\eqref{eq:totalvdw}  is written at zero temperature since finite-temperature
 effects in the VDW energy expression are negligibly small in the range of distances studied (as shown in
 Ref.~\cite{wetting}(Supplementary Material)). Of course the various atom-related characteristics have 
 to be used in the temperature regime where the liquid phase  is stable, as in 
 Section \ref{subsec:parameters}.

\begin{figure}[t]
\begin{center}
\includegraphics[width=0.95\columnwidth]{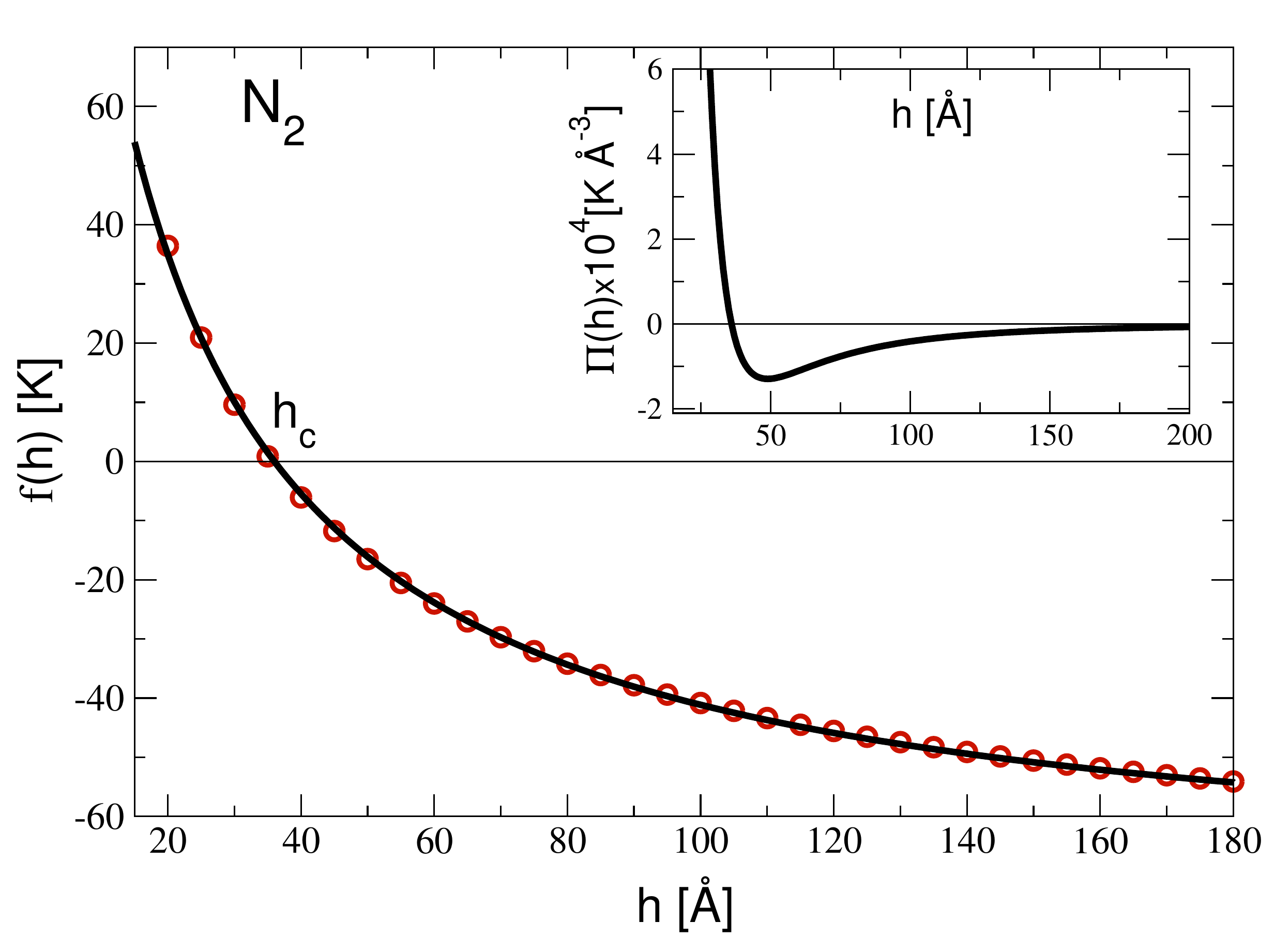}
\end{center}
\caption{The function $f(h)$, defined so that the disjoining pressure has the form $\Pi(h)=f(h)/h^3$. 
For N$_2$, $f(h)$ is calculated by evaluating the microscopic expressions Eqs.~\eqref{eq:totalvdw},\eqref{eq:dpdefinition},
and is shown as a solid line in the main panel.
The red circles represent a fit to the form $f(h) = -|\Pi_0| + \frac{\Pi_1}{h+L}$ (as in Eq.~\eqref{eq:dp}), with
$|\Pi_0|= 72.8 \ \text{K}, \  \Pi_1 = 3592  \ \text{K\AA},  \
   L = 13.3 \ \text{\AA}$.  These are the values used in the main text, Eq.~\eqref{eq:potN2}. Inset: The full function $\Pi(h)=f(h)/h^3$.
}
\label{fig:A1}
\end{figure}

\begin{figure}[h]
\begin{center}
\includegraphics[width=0.95\columnwidth]{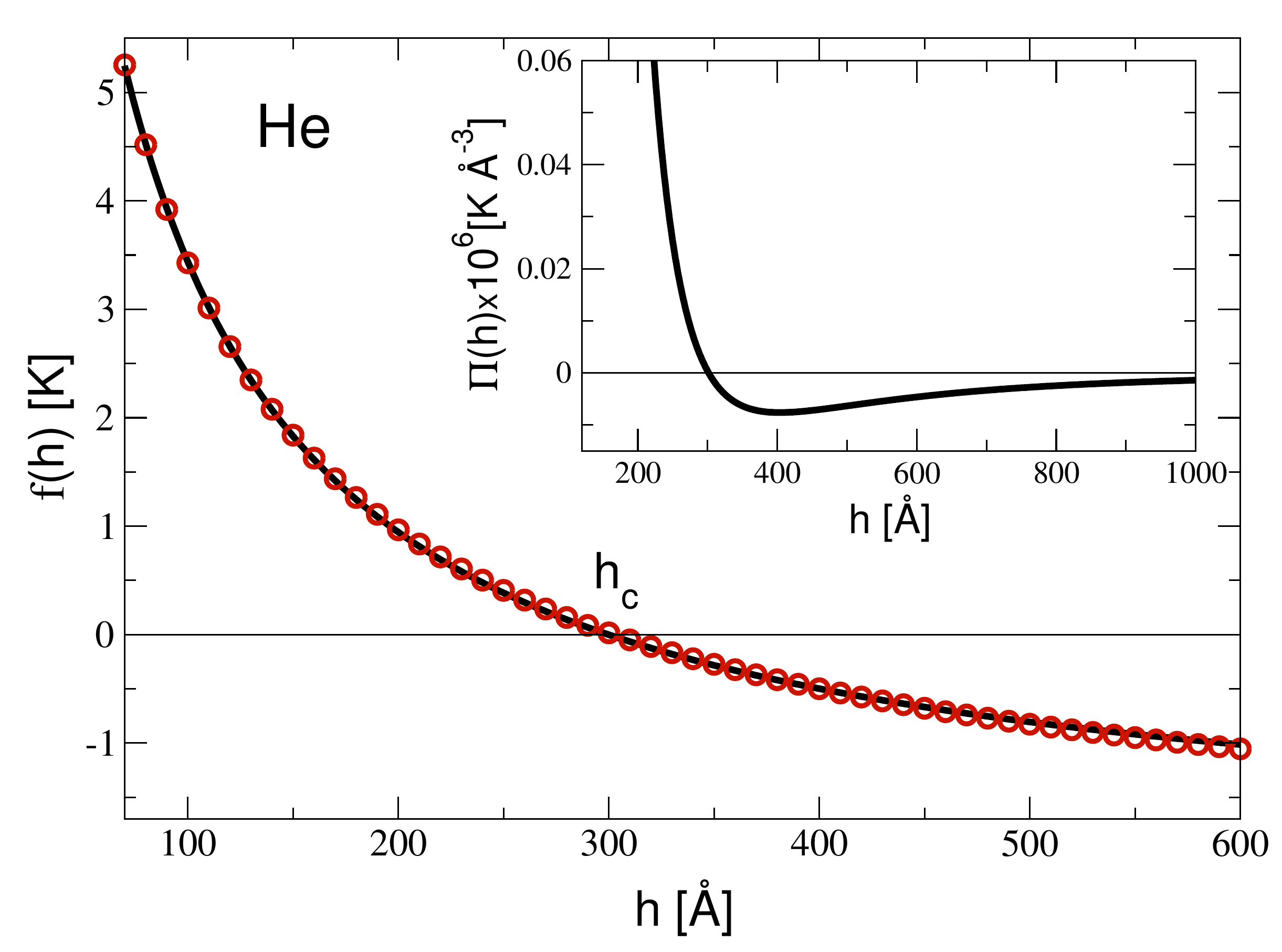}
\end{center}
\caption{ Results for He, following the same procedure as in Fig.~\ref{fig:A1}.
With $\Pi(h)=f(h)/h^3$, the main panel shows the exact numerical evaluation of $f(h)$, solid line.
The red circles represent a fit to the form 
$f(h) = -|\Pi_0| + \frac{\Pi_1}{h+L}$ (as in Eq.~\eqref{eq:dp}), with $|\Pi_0|= 2.09   \ \text{K},  \  \Pi_1 = 676   \ \text{K\AA}, \  L =  22.1  \ \text{\AA}$. 
These are the values used in the main text, Eq.~\eqref{eq:potHe}. Inset: The full function $\Pi(h)=f(h)/h^3$. Notice the different scales on this graph compared to
Fig.~\ref{fig:A1}.
}
\label{fig:A2}
\end{figure}

Several additional comments are in order. First, the fact that  $U_{\mbox{vdw}}(h)$ involves integration over the imaginary frequency axis
is a common mathematical feature when writing the ground state energy of the system \cite{Israelachvili, Dzyaloshinskii:1961vc,LL9}. Second, notice that 
${\cal{U}}_d (q,i\omega)  < 0$, while ${\cal{U}}_g(q,i\omega)  > 0$ (since we always have $\varepsilon > 1$ , which reflects screening). 
This will be important in what follows. Third, the terms ${\cal{U}}_d$ and ${\cal{U}}_g$  depend on $h$ only through the exponential factor. The nontrivial dependence of 
 $U_{\mbox{vdw}}(h)$ on $h$ arises after integration over the momentum ${\bf q}$.
Notice also that graphene's polarization $\chi(q, i\omega) $   has a pronounced momentum dependence 
which reflects the motion of graphene's quasiparticles. 

For completeness we also summarize the dielectric functions of the three liquids used in this work,
as described in \cite{wetting}, which cites additional literature.
  For Helium  the dynamical
dielectric constant is 
\begin{equation}
 \varepsilon(i\omega) = 1 +4\pi n_{\rm He} \alpha(i\omega),  \  \
\alpha(i\omega) = \frac{\alpha_{\rm He}}{1+ (\omega/\omega_{\rm He})^2}\, , 
\label{eq:Hepolarizability}
\end{equation}
where the density $n_{\rm He} = \SI{2.12E-2}{\angstrom^{-3}}$, the static
polarizability  $\alpha_{\rm He} = 1.38$~a.u., and the characteristic oscillator
frequency $\omega_{\rm He} = 27.2$ eV.  The atomic unit of polarizability is
defined as $1~\text{a.u.} = \SI{0.148}{\angstrom^3}$.
For Nitrogen and Hydrogen, which have  densities  comparable to Helium but significantly larger polarizabilities, more accurate
formulas based on the Clausius--Mossotti relation are typically used:
\begin{equation}
 \varepsilon(i\omega) = 1 + \frac{4\pi n_A \alpha(i\omega)}{1-
 \frac{4\pi}{3} n_A \alpha(i\omega)},  \   \   A ={\mbox{N}}_2, {\mbox{H}}_2\,
 \label{eq:NHpolarizability}
  \end{equation}
The dynamical polarizability $\alpha(i\omega)$ is defined as in Eq.~(\ref{eq:Hepolarizability}),
i.e. has the form  $\alpha(i\omega) = \frac{\alpha_{\rm A}}{1+ (\omega/\omega_{\rm A})^2}$. For
${\mbox{H}}_2$ the parameters are: $n_{\rm H_{2}} = \SI{2.04E-2}{\angstrom^{-3}}$,
$\alpha_{\rm H_2} = 5.44$~a.u., $\omega_{\rm H_{2}} = \SI{14.09}{\eV}$.  For
${\mbox{N}}_2$: $n_{\rm N_{2}} = \SI{1.73E-2}{\angstrom^{-3}}$, $\alpha_{\rm N_2}
= 11.74$~a.u., $\omega_{\rm N_{2}} = \SI{19.32}{\eV}$.

The VDW energy defined in Eq.~\eqref{eq:totalvdw} has physical dimensions of energy per unit area. The disjoining 
pressure is defined as:
\begin{equation}
\Pi(h) = - \frac{\partial  U_{\mbox{vdw}}(h)}{\partial h},
\label{eq:dpdefinition}
\end{equation}
and describes the effective force per unit area between the two boundaries of the system (liquid--vapor and liquid--graphene). It is clear that the part of $\Pi(h) $ which comes from ${\cal{U}}_g(q,i\omega)>0$ leads to positive
pressure, i.e. favors film growth, while the part associated with ${\cal{U}}_d (q,i\omega)  < 0$ is always negative, 
i.e. favors an instability. It is the competition between these two terms that leads to the spinodal de-wetting instability phenomenon.

Finally we return to the way we determine  the all-important functional form of  $\Pi(h)$, Eq.~\eqref{eq:dp}, 
which follows from the microscopic expressions Eqs.~\eqref{eq:totalvdw},\eqref{eq:dpdefinition}.
First we present the following qualitative considerations. As mentioned previously, it is
useful to consider the contributions of  the ${\cal{U}}_{d,g} (q,i\omega)$ terms separately. The (attractive) ${\cal{U}}_{d}$
part clearly leads to dependence of the form $\Pi(h) \sim -\frac{1}{h^3}$ which follows from counting powers of momenta in the integrals. The (repulsive) ${\cal{U}}_{g}$ part, however, exhibits a higher power due to the presence
of graphene's polarization $\chi(q, i\omega) $.  Since at intermediate frequencies, which are dominant in the integration, the dependence of $\chi(q, i\omega) $ on momentum
is quadratic for low momenta, this leads to  $\Pi(h) \sim {1}/{h^4}$ .
The exact way this crossover happens has to be determined numerically, by evaluating the expression Eqs.~\eqref{eq:totalvdw},\eqref{eq:dpdefinition}, which can be done with high accuracy. In Fig.~\ref{fig:A1}
we show the way this procedure works for N$_2$ and, as another example,  in Fig.~\ref{fig:A2} we present the results for He. Most importantly, we can conclude that the 
functional form of $\Pi(h)$, Eq.~\eqref{eq:dp}, used in the main text, is very accurate.

\bibliographystyle{apsrev4-2}
\bibliography{refs}

\end{document}